# Ultrafast ratchet dynamics of skyrmion by defect engineering under gigahertz magnetic fields

Weijin Chen,[1,2,3] Linjie Liu,[2,3] and Yue Zheng[2,3]⋆

[1]*School of Materials, Sun Yat-sen University, 510275, Guangzhou, China*
[2]*State Key Laboratory of Optoelectronic Materials and Technologies, Sun Yat-sen University, 510275, Guangzhou, China*
[3]*Micro&Nano Physics and Mechanics Research Laboratory, School of Physics, Sun Yat-sen University, 510275, Guangzhou, China*

The novel ratchet motion of magnetic skyrmions driven by microwave magnetic fields, with the motion direction and speed tunable by field parameters, provides a promising route to drive magnetic skyrmions in materials with poor conductivity. However, as an indirect motion, skyrmion ratchet motion speed is generally low in comparison with the direct motions driven by current. Toward practical applications, it is important to ask if there are mechanisms to realize ultrafast ratchet motion of magnetic skyrmions and how such novel motion can be integrated into racetrack-type skyrmion devices. In this work, we explore the effects of defects and edges on the ratchet motion of magnetic skyrmions under time-varying magnetic fields in GHz. We demonstrate that the ratchet motion of skyrmion is not only guided along the defect tracks or edges, but also with a remarkable speed-up (with a factor over ten) compared with that in the bulk region. The skyrmion ratchet motion speed reaches 100 m/s along a straight defect track/edge and $10^9$ rad/s along a circular edge under a field of ~50 mT, comparable to those direct motions driven by currents. Moreover, the skyrmion ratchet motion along the defect track/edge can be facilely controlled by the field and defect parameters. Analysis based on time-averaged Thiele's equation of skyrmion verifies that such a speed-up effect is due to the increased time-averaged driving force perpendicular to the skyrmion motion when it approaches the defect track/edge, analogous to that discovered in direct motions driven by currents. The generality of our conclusions has been examined for the ratchet motion of Bloch and Néel-type skyrmions driven by a variety of time-varying magnetic fields, and for systems with open edges or defect tracks with modified Dzyaloshinskii–Moriya or exchange interactions and anisotropy.

⋆ e-mail: zhengy35@mail.sysu.edu.cn.

Magnetic skyrmions are a kind of particle-like chiral spin textures with a nontrivial topological feature (*i.e.*, cannot be continuously 'unwound' to a ferromagnetic or other magnetic states) and a typical size in range of $10^0 \sim 10^2$ nm[1,2]. They have been identified in both conductive and poorly conductive materials possessing Dzyaloshinskii–Moriya (DM) interaction, caused by the breaking of inversion symmetry in crystal lattice[3-8] or nearby the interfaces[9-15]. The relatively small size, robust metastability, as well as the intriguing dynamics and transport properties tunable by various external sources, make magnetic skyrmions appealing building blocks in ultradense information storing and processing devices. Particularly, the concepts of racetrack-type devices are improved by skyrmions, and the threshold current to displace magnetic skyrmions is extremely low compared with magnetic domain walls, implying the potential of developing highly energy-efficient devices.

An in-depth understanding of skyrmion motion characteristics is crucial for their promising applications. It is well-known fact that magnetic skyrmions can be displaced, e.g., by spin polarized currents via the spin-transfer torque (STT) or spin-orbit torque (SOT) applicable for metallic systems[16-18], and by field or field gradients in the absence of spin torques from the conduction electrons applicable for both metallic and insulating sytems[19-27]. The nontrivial topology of skyrmions manifests itself in the motion known as skyrmion Hall effect (SkHE)[28,29], a reverse effect of the topology-induced emergent fields acting on electrons and magnons. The SkHE would drive skyrmions to the edges of system and causes possible annihilation, limiting the maximum velocity of skyrmion. Meanwhile, when skyrmion is moving along the edges, a speed-up effect occurs due to the repulsive force of the edge[30,31]. So far, a lot of efforts have been paid to pursue high-speed and controllable motion of skyrmion. Particularly, researchers have proposed the use of racetrack-type structures (e.g., narrow stripes with natural edges or defect tracks with different anisotropy, DM or exchange interactions from those of the surroundings) to guide the motion of skyrmion[17,31-34]. With the further use of antiferromagnetic bilayer structures[35], low-symmetry heavy-metal substrates[36] or ferrimagnetic materials[37], the SkHE can be largely suppressed and the speed limit in skyrmion racetrack devices can be overcome by a factor of ten[36].

While the academic focus has been put on the direct motion of skyrmion driven by time-unvarying sources, some works have shown that a net motion of skyrmion can be also realized by indirect motion under time-varying sources[24-26,38-40]. Such motions, occurring in systems with space or time asymmetries, are relevant to a general kind of transport phenomena of solitons, named ratchet effect. Remarkably, both the speed and the direction of the skyrmion ratchet motions under time-varying magnetic fields can be tuned by the field parameters[25,40], indicating a general and feasible strategy to drive skyrmions in both conductive and poorly conductive materials. Nevertheless, being an indirect motion, skyrmion ratchet motion speed is generally low (<10 m/s) compared with the direct motions driven by current, despite relatively fast ratchet motion can be realized via a tradeoff of stability for speed (e.g., by increasing the driving field strength or setting the driving field frequency close to the resonance frequency of skyrmion[25,40]). Toward practical applications, it is important to ask if there are mechanisms to realize ultrafast ratchet motion of skyrmions and how such novel motion can be integrated into racetrack-type skyrmion devices.

Based on massive micromagnetic simulations and analytical modeling, this work explores the effects of defects and edges on the ratchet motion of magnetic skyrmions driven by gigahertz magnetic fields and demonstrates a general mechanism of ultrafast ratchet motion (over 100m/s under a field of ~50 mT) of magnetic skyrmions via defect engineering. The physical model and the basic idea of defect-induced ultrafast ratchet motion of skyrmion are schematically illustrated in Fig. 1. We consider a novel ratchet motion of skyrmion driven by GHz magnetic fields with a temporal asymmetry[40], but our conclusions should not be limited to such a ratchet motion. Two types of driving fields with a temporal asymmetry are focused: in-plane biharmonic magnetic fields $H_x(t) = H_0[\sin(m\omega t) + \sin(n\omega t + \varphi)]$ with $m$ and $n$ being two coprime integers and $m + n$ being odd, and in-plane pulsed magnetic fields with positive pulses and negative pulses in different magnitudes or widths. Due to the temporal asymmetry of the driving fields, skyrmion can be driven into a net motion even though the time average of driving fields is zero and the system is without any space asymmetry. In the bulk region where the skyrmion is far away from defect tracks or edges, the skyrmion would show ratchet motions with a relatively low speed compared with that near the defect tracks or edges[40]. When the skyrmion approaches to a defect track or an edge, it will be guided along the defect track/edge and the skyrmion ratchet motion exhibits a much larger speed than that in the bulk region. Analogous to the mechanism discovered in direct motions driven by currents[30,34], the speed-up effect of skyrmion ratchet motion can be explained by the repulsive interactions (in a time-averaging view) between the skyrmion and the defect track/edge. We will show that the skyrmion ratchet motion speed and direction along the defect track/edge can be facilely controlled by the field and defect parameters. Our conclusions are general for the ratchet motion of Bloch and Néel-type skyrmions driven by time-varying magnetic fields in systems with open edges or defect tracks (*i.e.*, with modified Dzyaloshinskii–Moriya or exchange interactions and anisotropy).

**Results**

**Defect-mediated skyrmion ratchet motion under time-asymmetric magnetic fields.**
The ratchet effect under time-asymmetric fields is general for both Bloch and Néel-type magnetic skyrmions as they are identical particles from the viewpoint of topology (see Supplementary Fig. S1 and S2). We would take the results of Bloch-type skyrmions to illustrate the key idea of this work. Our simulations are performed on a two-dimensional (2D) Heisenberg-type spin lattice with 128 × 128 sites (see Methods Section for details of the model and values of the model parameters). The temperature-field $T$-$H_z$ phase diagram of single-skyrmion state of this spin lattice system is calculated and plotted in Supplementary Fig. S3. It shows that, at a given temperature, the single-skyrmion state (*i.e.*, an isolating skyrmion embedded in a ferromagnetic background) can be stabilized at an intermediate range of the out-of-plane magnetic field $H_z$, consistent with previous works[21,22]. In the following, we fixed the condition at $T = 0$ K and $H_z = 0.015$ (in unit of $J/g\mu_B$, see Methods Section). Such a condition stabilizes a skyrmion with diameter of ~60 nm if we consider a typical lattice parameter of $a = 1$ nm. A finite change of temperature and $H_z$ would not change the main conclusions of this work.

We first reveal the defect-mediated ratchet motion of skyrmion under a biharmonic magnetic field $H_x(t) = 0.003[\sin(\omega t) + \sin(3\omega t)]$ along $x$ axis. The angular frequency of the fields $\omega$ is set to be 105 rad/s (also in the following). The corresponding frequency is $f = 16.7$ GHz. This value is close to the resonance frequency of the counter-clockwise (CCW) gyration mode of the skyrmion and leads to a relatively fast ratchet motion[39]. Two cases are investigated. For the first case, denoted as "defect track // $y$", the $128a \times 128a$ spin lattice system is in periodic boundary conditions along both $x$ and $y$ axes, and a $4a$-wide defect track (with the DM interaction constant being half of that in the bulk region, *i.e.*, $\eta_{DMI} = 0.5$) parallel to $y$-axis is located at $x = 96a$. For the second case, denoted as "edge // $x$", the spin lattice system has a periodic boundary condition along $x$-axis and an open-circuit condition along $y$-axis (*i.e.*, there are two edges parallel to $x$-axis). For both cases, an isolating skyrmion is initially located at the center of the system $(64a, 64a)$. The core of skyrmion before motion is far enough from the defect track and edges, therefore the skyrmion motion at the beginning can be considered as that in the bulk region. Figure 2a and 2b respectively depict the motion trajectories of the skyrmion in the two cases. It is clearly seen that for both cases the skyrmion exhibits a ratchet motion with a helical trajectory under the biharmonic in-plane magnetic field. At the beginning, the pitches of the helical trajectories of skyrmion for both cases are rather small (see the "1#" insets in Fig. 2a and 2b), indicating a small ratchet speed. As the skyrmion moves closer to the defect track or the edge, the pitches of the helical trajectories of skyrmion for both cases gradually increase, and become much larger than that in the bulk region when the skyrmion finally moves along the defect track or the edge (see the "2#" insets in Fig. 2a and 2b). This clearly demonstrates a speed-up effect of the skyrmion ratchet motion caused by the defect track and edge, analogous to those reported in the direct transport of skyrmion driven by currents[30,34].

Such a speed-up effect can be quantitatively inferred from the long-time evolution curves of the skyrmion position coordinates $x_c$ and $y_c$ of the two cases, as shown in Fig. 2c and 2d, respectively. It shows that the $y_c$ curve of case "defect track // $y$" and the $x_c$ curve of case "edge // $x$" have smooth slopes at the beginning of several nanoseconds (the unit of time $\tau$ is about 1.5 ns, see Methods Section). This corresponds to the stage of skyrmion approaching the defect track or the edge. Then, the curves show periodic zigzags with steep slopes, corresponding to the stage of high-speed ratchet motion along the defect track or the edge. Note, due to the use of periodic boundary conditions, each zigzag of the $y_c$ or $x_c$ curve corresponds to a 128$a$ displacement of the skyrmion, and the period of the zigzag is the time for the skyrmion to move over 128 lattices. One can readily estimate the ratchet motion speed $v_c$ along the defect track and the edge to be about 68.7$a/\tau$ and 147.9$a/\tau$, respectively. For a typical value of $a = 1$ nm, the speed values are 45.8 m/s and 98.6 m/s, respectively, which are comparable to that driven by currents. On the other hand, from the slope of the $y_c$ or $x_c$ curve at the beginning (see, *e.g.*, in the "1'#" inset in Fig. 2c), the ratchet motion speed in the bulk region is about 9.54 $a/\tau$ (*i.e.*, 6.4 m/s for $a = 1$ nm). This result shows that the speed-up effect of defect track and edge on the skyrmion ratchet motion is remarkable, with a factor over 10.

To clearly see the change of skyrmion configuration under the biharmonic in-plane field and the effect of defect, in Fig. 2e we depict the snapshots of the contour map of

$m_z$ during the motion of skyrmion towards and along the edge (case "edge // $x$"). The movies of skyrmion ratchet motion along the defect track ("defect track // $y$") and the edge ("edge // $x$") are further provided in the Supplementary Information. One can see from these results that, under the driving force of the biharmonic field, the skyrmion is excited into a CCW gyration mode with a significant time-varying deformation. When the skyrmion approaches the edge or defect track, it shows a slight shrinkage along the direction perpendicular to the edge or defect track. Such a skyrmion-defect interaction reflects that the skyrmion is subjected to a repulsive force from the edge or defect track, in addition to the driving force of the biharmonic magnetic field. As we will show later, this repulsive force is crucial for the speed-up effect of skyrmion ratchet motion.

**Facile control on the defect-induced ultrafast ratchet motion of skyrmion.** Note, by varying the phase $\varphi$ of the biharmonic field in form of $H_x(t) = h_1\sin(m\omega t) + h_2\sin(n\omega t + \varphi)$, direction of the skyrmion ratchet motion in the bulk region (*i.e.*, where skyrmion is far away enough from defect tracks or edges) can be continuously rotated by 360° (Supplementary Fig. S4). Consequently, the skyrmion will be driven towards/along or away from a defect track/edge, depending on the phase $\varphi$ of the biharmonic field. This 360° ratchet motion of skyrmion controllable by the phase $\varphi$ is an analogy to the 360° direct motion of skyrmion by controlling the magnetization direction of the spin current injected in the out-of-plane direction[18], and it is potentially useful in driving skyrmions in insulating systems. Double-track or double-edge structures can be further designed to confine the skyrmion ratchet motion within the region between double defect tracks or double edges, acting like racetrack-type highways of skyrmion ratchet motion. In the following, we discuss the influence of phase $\varphi$ on the skyrmion ratchet motion within a double-track structure under biharmonic in-plane magnetic fields $H_x(t) = 0.003[\sin(\omega t) + \sin(3\omega t + \varphi)]$ with $\varphi$ varying from 0 to 360°. In the simulated $128a \times 128a$ spin lattice system, two $4a$-wide defect tracks (parallel to the $y$-axis, $\eta_{DMI} = 0.5$) are located at $x = 32a$ and $x = 96a$. For a given $\varphi$, an isolating skyrmion is initially located at position ($64a$, $64a$).

The long-time evolution curves of the position coordinate $y_c$ and the corresponding trajectories of skyrmion ratchet motion at various phase $\varphi$ are depicted in Fig. 3a and 3b, respectively. One can see from the number of the zigzags in the $y_c$ curves in Fig. 3a that the ratchet motion speed of skyrmion along the defect track is sensitive to phase $\varphi$. In addition, although the direction of the skyrmion ratchet motion in the bulk region is 360° rotatable (Supplementary Fig. S4), the skyrmion is finally either in upward ratchet motion along the left defect track // $y$ or in downward ratchet motion along the right defect track // $y$ as shown in Fig. 3b. That is to say, the rachet motion of skyrmion along the defect track has a fixed direction (which can be represent by $\mathbf{e}_z \times \mathbf{e}_d$, with $\mathbf{e}_z$ being the vector along out-of-plane direction and $\mathbf{e}_d$ being the vector perpendicular to the defect track and pointing to the skyrmion), no matter the skyrmion has an initial upward or downward velocity when it approaches the defect track. Again, this indicates that the repulsive force of the defect track (which is fixed in direction $\mathbf{e}_d$) plays a dominant role in the ratchet motion of skyrmion.

The skyrmion ratchet motion speed and the speed-up factor as a function of phase

$\varphi$ are calculated and depicted in Fig. 3c, both for motion along the defect track and in the bulk region. Here, the speed along the defect track $v$ is plotted with plus or minus sign, denoting the upward motion along the left defect track or downward motion along the right defect track. Moreover, the speed-up factor is defined as the ratio between the ratchet motion speed along the defect track (determined by the slope of the zigzag) and that in the bulk region (determined by the slope of the $y_c$ curves at the beginning 1 ns). One can see that both the ratchet motions in the bulk region and along the defect track are tunable by phase $\varphi$. The $v$-$\varphi$ curves are sinusoidal-like, with a near 40° difference between the peaks of the ratchet motion speed along the defect tracks and that of the bulk region. The maximum ratchet motion speed along the defect tracks is about 91.3$a/\tau$ at $\varphi$ around 130° (downward along the right track) and 310° (upward along the left track), whereas the maximum ratchet motion speed in the bulk region is about 9.5 $a/\tau$ at $\varphi$ around 0° and 180°. The maximum speed-up factor is over 10 at $\varphi$ around 105° and 285°, and the minimum speed-up factor is about 2.2 at $\varphi$ around 35° and 215°.

The $\varphi$-dependent speed-up effect of the skyrmion ratchet motion along the defect track can be further modified by the defect parameters. Figure 3d depicts the skyrmion ratchet speed $v_y$ as a function of phase $\varphi$ of the biharmonic field for cases with different defect strength $\eta_{DMI}$. Effect of the width of defect track $w$ is also shown in the inset. It can be seen that the ratchet motion speed is more sensitive to the defect strength than the width of defect track, implying the dominant effect comes from the border of the defect track. The defect strength mainly affects the ratchet motion speed rather than shifting the $v$-$\varphi$ curve along the $\varphi$ axis. With the change of the defect strength $\eta_{DMI}$ from

0.8 to 0.5, the maximum ratchet motion speed increases from about $68.5a/\tau$ to $91.3a/\tau$ (by a factor of ~1.3). Moreover, depinning occurs when the defect strength is not strong enough ($\eta_{DMI} = 0.8$), indicating that the effective driving force exerted to the skyrmion exceeds the depinning force of the defect track. The change of the width of defect track from $w = 4a$ to $12a$ slightly shifts the $v$-$\varphi$ curve by ~15°, with a minor effect on the maximum speed.

To clearly see the effect of phase $\varphi$ on the skyrmion ratchet motion along the defect track, we depict the polar graphs of the skyrmion ratchet motion speed and the speed-up factor as a function of phase $\varphi$ (Fig. 3e and 3f). The polar curves of ratchet motion speed are in peanut-shape with two peaks and valleys. The polar curve of ratchet motion speed along the defect tracks have a near 40° rotation with those of the bulk region (Fig. 3e). As a result, the polar curve of the speed-up factor shows a distortion from that of the speed curve (Fig. 3f). Note, the motion angle $\theta$, defined as the angle between the velocity vector and the $x$-axis, is 360° rotatable with the change of phase $\varphi$ when the skyrmion moves in the bulk region (Fig. 3c). By plotting the polar graphs of the ratchet motion speed and the speed-up factor as a function of the motion angle $\theta$ in the bulk region (Supplementary Fig. S5), one can see that the fastest ratchet motion along the defect track actually occurs when $\theta$ is about 0 and 180°, that is, when the skyrmion is perpendicularly incident to the defect track. For this motion angle, the repulsive force of the defect track should be maximum, again implying that the important role of the repulsive force of the defect track in the ultrafast ratchet effect.

Skyrmion ratchet motion along the edge driven by biharmonic in-plane magnetic fields is found to be much less sensitive to phase $\varphi$ than the case of motion along defect track (see Supplementary Fig. S6 and S7). In the simulation, there are two open edges at the two boundaries of the lattice systems parallel to *x*-axis as that shown in Fig. 2b. With phase $\varphi$ changing from 0° to 360°, the direction of the skyrmion ratchet motion is either in rightward motion along the upper edge or in leftward motion along the bottom edge. Moreover, the ratchet motion speed along the edge is generally large, which is about 150 $a/\tau$ for all $\varphi$ and is about 1.7 times of the maximum speed moving along the defect track (~90 $a/\tau$). The even faster ratchet motion along the edge indicates a stronger interaction between the skyrmion and the edge, and the much weaker phase-sensitivity in contrast to the case of defect track is believed to be caused by much more intensive excitation of spin waves nearby the edge, with an intensity insensitive to phase $\varphi$. For this case, it would be more efficient to tune the ratchet motion speed along the edge by controlling the field strength of the biharmonic field, or by controlling the pulse width of pulsed magnetic fields as shown in Fig. 4. Here, the pulsed fields $H_x(t)$ are repetitive sequences of alternating positive pulse over a time $TN_1$ and negative pulse over a time $TN_2$ (Fig. 4a). The field strength of the positive and negative pulses is 0.003 in reduced unit (~33 mT), and $TN_1 + TN_2 = TN$ is fixed to be $0.04\tau$ (~60 ps). From the long-time evolution curves of the position coordinate $x_c$ of skyrmion ratchet motion under pulsed fields with different $TN_1$ (Fig. 4b), one can see that the skyrmion ratchet motion speed along the edge significantly increases with the change of $TN_1$ from $0.002\tau$ to $0.01\tau$,. The facile control of the ratchet motion speed by pulsed fields is believed to be caused by the resonance effect and is practically useful for driving skyrmion in nanotracks with open edges.

**Mechanism analysis of the defect-induced ultrafast ratchet motion of skyrmion.**
To understand the origin of the defect-induced ultrafast ratchet motion of skyrmion, we further make an analysis on the skyrmion motion based on the time-averaged Thiele's equation. It should be pointed out that, under the effect of time-varying magnetic fields, the skyrmion system is excited with fast spin modes (magnons) and the skyrmion shows significant dynamic deformations. This drives the skyrmion transport beyond the rigid-particle approximation, and it is necessary to consider the skyrmion-magnon interaction. Here, making use of the periodic property of the driving magnetic field and the fact that the spin modes relax much faster than the translation mode, one can adopt a similar idea of previous works[21,24] to derive a time-averaged Thiele's equation for the ratchet motion of skyrmion (see Methods Section for the details). The time-averaged Thiele's equation is still in classic form as **G**×**v** + **Dv** = **F**, where **v** and **F** are the time-averaged velocity and effective driving force over a period, and **G** and **D** being the effective gyromagnetic vector and the dissipative force tensor determined by the time-averaged magnetization over a period, respectively. The effective driving force **F** includes the contributions of the external field, the interaction force exerted by the defect track or the edge, as well as the skyrmion-magnon interaction (see Eq. (16) in Methods Section).

Figure 5a depicts the snapshots of the force density map when a skyrmion moves toward and along an edge parallel to $x$-axis under a biharmonic in-plane magnetic field $H_x(t) = 0.003[\sin(\omega t) + \sin(3\omega t)]$ with $\omega = 105$ rad/s. The simulation setting is the same as that in Fig. 2. One can see that the force density takes large values mainly near the skyrmion. When the skyrmion approaches the edge, the force changes direction from being lower-right to being perpendicular to the edge and shows a remarkable increase in magnitude as indicated by the ranges of the color bars. To see the difference between the driving forces subjected by the skyrmion when it moves in the bulk region and along the defect track or the edge, we further depict in Fig. 5b the evolution curves of time-averaged driving force components $F_{x(y)}$ subjected by skyrmion under biharmonic field $H_x(t) = 0.003[\sin(\omega t) + \sin(3\omega t)]$ with $\omega = 105$ rad/s. For the cases of "defect track // $y$" and "edge // $x$", the simulation settings are the same as those in Fig. 2. Our result shows that the driving force is small for ratchet motion without defect tracks or edges (case "bulk") or far from the defect tracks or edges (cases "defect track // $y$" and "edge // $x$" at the beginning). Moreover, the driving force remarkably increases as the skyrmion approaches the defect track or edge. At steady motion, the driving force is found to be (in unit of $a/\tau$) $\mathbf{F} = (F_x, F_y) = (-72.6, -62.5)$ for case "bulk", $(-545.6, -49.4)$ for case "defect track // $y$", and $(104.4, -1367.3)$ for case "edge // $x$", respectively. According to these values, we estimate the ratchet motion speed $v$ and motion angle $\theta$ of the three cases based on the approximated relation $\mathbf{G} \times \mathbf{v} \sim \mathbf{F}$ and $\mathbf{G} \sim -4\pi \mathbf{e}_z$. This gives $v = 7.6a/\tau$ and $\theta = 310.7°$ for case "bulk", $v = 43.6a/\tau$ and $\theta = 275.2°$ for case "defect track // $y$", and $v = 109.1a/\tau$ and $\theta = 4.4°$ for case "edge // $x$". These estimated values are in good consistence with the simulation results, verifying the correctness of our physical picture. Therefore, it can be concluded that the speed-up effect of skyrmion ratchet motion by defect track or edge is caused by the remarkable increase of the effective driving force perpendicular to the skyrmion motion when it approaches the defect track or the edge (Fig. 5c).

Note, there are actually various situations when a skyrmion encounters the defect track/edge as schematically shown in Fig. 5d. Depending on the relative magnitudes of the force perpendicular to the defect track/edge $F$ subjected by skyrmion and the critical force $F_c$ of depinning or annihilation, three situations of skyrmion motion can happen: (i) skyrmion moves along the defect track or the edge ($F < F_c$), (ii) passes through the defect track ($F > F_c$ of depinning), and (iii) annihilates at the defect track or the edge ($F > F_c$ of annihilation). The force of a defect track or an edge exerted to the skyrmion undergoing ratchet motion depends on the distance between the skyrmion and the defect track/edge and the defect parameters (*e.g.*, the defect strength and track width), similar to the cases of interactions between skyrmion and defect track/edge without significant excitation[41]. However, for our case, the skyrmion-magnon interaction also contributes to the force of a defect track or an edge exerted to the skyrmion (see Eq. (16) in Methods Section). $F_c$ of depinning should be also a function of the defect parameters, whereas $F_c$ of annihilation is largely affected by the skyrmion stability[42].

Based on the above force analysis, we can further explain the controllability of the speed-up effect along the defect track or edge by the field and defect parameters. Take the skyrmion ratchet motion along defect track // $y$ and along edge // $x$ under biharmonic fields $H_x(t) = 0.003[\sin(\omega t) + \sin(3\omega t + \varphi)]$ as an example. In previous results, we have shown that the ratchet motion speed along the defect track is highly tunable by phase $\varphi$ (Fig. 3), meanwhile the ratchet motion along the edge is much less sensitive to phase $\varphi$

(Supplementary Fig. S6 and S7). For these two cases, the different $\varphi$-sensitivity of the ratchet motion speed is due to the different $\varphi$-dependences of the time-averaged driving force **F** as shown in Fig. 5e. One can see that the force curves of the two cases are highly consistent with the speed curves (Fig. 3c and Supplementary S7). The force curves of case "defect track // *y*" is much smoother than that of the case "edge // *x*". Moreover, it is clearly seen from the force curves of case "defect track // *y*" that the peak and valley values of the rachet motion speed correspond to the maximum and minimum values of $F_x$. Figure 5f further shows a plot of the skyrmion ratchet speed as a function of driving force based on the data of Figure 5e. A clear linear correlation between speed and force is clearly seen from the data of "defect track // *y*", in consistence with the approximated relation $\mathbf{G} \times \mathbf{v} \sim \mathbf{F}$. From the slopes of $v_{x(y)}$ vs. $F_{x(y)}$ curves, one can estimate the magnitude of the "effective" gyromagnetic vector **G** is about 8.03 and 10.13 for the cases of "defect track // *y*" and "edge // *x*", respectively. These values are smaller than $4\pi$, implying the difference between the time-averaged magnetization skyrmion during ratchet motion and that at the equilibrium.

**Discussion**

The ultrafast ratchet motions can be realized along defect tracks which have a different exchange interaction or anisotropy from that of the bulk region (Supplementary Fig. S8 and Fig. S9). The ratchet motion along such defect tracks is also highly tunable by the field parameters and the defect strength. Moreover, an ultrafast rotation motion with the angular velocity being over $10^9$ rad/s of skyrmion can be realized when the skyrmion is confined within a circular disk (Supplementary Fig. S10). More complicate motions of skyrmion can be realized by proper defect engineering, geometrical designing, and control of the field parameters. The same mechanism is also applicable for Néel-type skyrmions (Supplementary Fig. S11). The reported ultrafast skyrmion ratchet motion is a conjoint effect of the soliton-like feature of magnetic skyrmions and skyrmion-defect interaction or the skyrmion-edge interaction. The driving sources can be magnetic fields or other time-varying sources with gigahertz characteristic frequencies rather than spin polarized currents. As shown by our results, magnetic fields of about 50 mT can trigger efficient ratchet motion along defect tracks or edges in a speed of about 100 m/s. Note, the presence of defects or edges in the skyrmion system lowers the spatial symmetry of the system, and ratchet motion of skyrmion along a defect track or edge can actually be induced by time-varying sources without time asymmetry[43,44]. The reported mechanism therefore provides a general and feasible strategy to efficiently drive skyrmion in both metallic and insulating systems based on racetrack structures. Our conclusions can be extended to other Magnus-dominated particle systems.

**Methods**

**Micromagnetic simulation.** Basing on a Heisenberg model on a 2D square lattice, we write the following effective Hamiltonian of a chiral magnet as

$$\mathcal{H}(\mathbf{m}_i) = -J \sum_{<i,j>} \mathbf{m}_i \cdot \mathbf{m}_j - D\left(\sum_i \mathbf{m}_i \times \mathbf{m}_{i+e_x} \cdot \hat{e}_x + \sum_i \mathbf{m}_i \times \mathbf{m}_{i+e_y} \cdot \hat{e}_y\right) \\ - \sum_i \mathbf{H}(t) \cdot \mathbf{m}_i - K m_z^2 \quad (1)$$

where $\mathbf{m}_i$ is the magnetization vector at site $I$, $J$ is the Heisenberg exchange coefficient,

$D$ is the DM interaction coefficient, $K$ is the anisotropy constant, and $\mathbf{H}(t) = \mathbf{H}_0 + \mathbf{h}(t)$ is the external magnetic field which is the sum of a constant field normal to the plane $\mathbf{H}_0 = (0, 0, H_z)$ and a time-varying in-plane field $\mathbf{h}(t)$.

The dynamics of skyrmion is captured by solving the stochastic Landau-Lifshitz-Gilbert (LLG) equation,

$$\frac{d\mathbf{m}_i}{dt} = -\gamma \left[ \mathbf{m}_i \times \left( \mathbf{H}_i^{\mathrm{eff}} + \mathbf{L}_i^{\mathrm{fl}}(t) \right) \right] + \alpha \left( \mathbf{m}_i \times \frac{d\mathbf{m}_i}{dt} \right) \quad (2)$$

or in the equivalent form,

$$\frac{d\mathbf{m}_i}{dt} = -\frac{\gamma}{\alpha^2 + 1} \left[ \mathbf{m}_i \times \left( \mathbf{H}_i^{\mathrm{eff}} + \mathbf{L}_i^{\mathrm{fl}}(t) \right) + \alpha \mathbf{m}_i \times \left[ \mathbf{m}_i \times \left( \mathbf{H}_i^{\mathrm{eff}} + \mathbf{L}_i^{\mathrm{fl}}(t) \right) \right] \right] \quad (3)$$

where $\gamma = g\mu_B / \hbar$ is the gyromagnetic ratio, $\alpha$ is the Gilbert damping coefficient, $\mathbf{H}_i^{\mathrm{eff}}$ is the effective magnetic field given by $\mathbf{H}_i^{\mathrm{eff}} = -\partial \mathcal{H} / \partial \mathbf{m}_i$, and $\mathbf{L}_i^{\mathrm{fl}}(t)$ is the stochastic field caused by the effects of a thermally fluctuating environment interacting with $\mathbf{m}_i$. $\mathbf{L}_i^{\mathrm{fl}}(t)$ satisfies $\langle \mathbf{L}_i^{\mathrm{fl}}(t) \rangle = 0$ and $\langle L_{i\beta}^{\mathrm{fl}}(t) L_{i\lambda}^{\mathrm{fl}}(s) \rangle = \alpha k_B T \gamma^{-1} m^{-1} \delta_{ij} \delta_{\beta\lambda} \delta(t - s)$, where $\beta$ and $\lambda$ are Cartesian indices, $T$ is temperature, $k_B$ is the Boltzmann constant, and $m = |\mathbf{m}_i| = |g\mu_B| / a^3$ is the norm of the magnetization vector.

Numerical simulations based on the stochastic LLG equation are performed via an explicit Euler iteration scheme. The size of sample systems is fixed to be $128 \times 128$ sites under the periodic boundary condition. In the bulk region, the Heisenberg exchange $J$ is taken to be $J/k_B = 50$ K, the strength of the DM interaction coefficient is $D = 0.15J$, and the anisotropy constant $K$ is taken to be zero. The spin turn angle $\theta$ in the helical structure is ~6° as determined by $\theta = \arctan[D / (\sqrt{2}J)]$. The Gilbert damping coefficient $\alpha$ is taken to be 0.1. The external magnetic field normal to the plane is fixed to be $H_z = 0.015$ in unit of $J / (g\mu_B)$, which is ~0.16 T for $g$ equal to 6.74. The time step is taken to be 0.01, in unit of $\hbar/J$, which is ~1.5 fs. In the main text, the time is expressed in the reduced form $\tilde{t}=t/\tau$, with $\tau$ being $10^4\hbar/J \simeq 1.5$ ns. To first obtain the steady skyrmion, the magnetic structure of the spin lattice system is initially set with a downward magnetization in the center region and with an upward magnetization elsewhere and is relaxed over a sufficiently long time (>1 ns).

To characterize the skyrmion, we calculate the topological charge density,

$$q = \frac{1}{4\pi}\mathbf{m}\cdot(\partial_x\mathbf{m}\times\partial_y\mathbf{m}) \tag{4}$$

as defined in the continuous form. The total topological charge is then given by

$$Q = \int q\,dxdy \tag{5}$$

And the position of skyrmion $\mathbf{R}=(x_c, y_c)$ can be determined by

$$x_c = \frac{\int xq\,dxdy}{Q}, \quad y_c = \frac{\int yq\,dxdy}{Q} \tag{6}$$

**Analytical model of skyrmion ratchet motion.** Consider the steady ratchet motion of skyrmion driven by a periodically oscillating magnetic field. In this case, spin waves (magnons) are excited and deforms the skyrmion dynamically, and the conventional Thiele's equation based on the rigid-particle model needs to be modified. Assume that the characteristic frequency of the driving field is in order of gigahertz, which is much lower than those of spin waves (magnons). The magnetization of a skyrmion is written as the sum of a slowly-varying part and fast-varying part[21,24],

$$\mathbf{m}(\mathbf{r},t) = \mathbf{m}_0 + \mathbf{n} \tag{7}$$

where $\mathbf{m}_0$ is the slowly-varying part and $\mathbf{n}$ is the fast-varying part related to the magnons, respectively. For simplicity, we make approximations on $\mathbf{m}_0$ and $\mathbf{n}$ that $\mathbf{m}_0=\langle\mathbf{m}(\mathbf{r},t)\rangle$,

$\langle \mathbf{n} \rangle = 0$, with $\langle\ \rangle$ denoting time averaging over a period. Hence, $\mathbf{m}_0$ is the time average of the skyrmion magnetization over a period. The time-dependence of $\mathbf{m}_0$ captures the net motion of skyrmion, and it is related to the time-average of the skyrmion position $\langle \mathbf{R}(t) \rangle$ and the time-averaged skyrmion velocity $\mathbf{v} = \langle \dot{\mathbf{R}} \rangle$ as $\mathbf{m}_0 = \mathbf{m}_0 (\mathbf{r} - \langle \mathbf{R}(t) \rangle)$ and $\dot{\mathbf{m}}_0 = -(\mathbf{v} \cdot \nabla) \mathbf{m}_0$.

Substituting the above $\mathbf{m}(\mathbf{r}, t)$ into the LLG equation (see, Eq. (2)), one obtains

$$-(\mathbf{v} \cdot \nabla) \mathbf{m}_0 + \dot{\mathbf{n}} = -\gamma \mathbf{m} \times \mathbf{H}_{\text{eff}} + \alpha (\mathbf{m}_0 + \mathbf{n}) \times \left( -(\mathbf{v} \cdot \nabla) \mathbf{m}_0 + \dot{\mathbf{n}} \right) \tag{8}$$

Multiplying $\mathbf{m}_0 \cdot \left( \dfrac{\partial \mathbf{m}_0}{\partial x_i} \times \right.$ on all the terms of the above equation gives

$$-\mathbf{m}_0 \cdot \left( \frac{\partial \mathbf{m}_0}{\partial x_i} \times (\mathbf{v} \cdot \nabla) \mathbf{m}_0 \right) + \mathbf{m}_0 \cdot \left( \frac{\partial \mathbf{m}_0}{\partial x_i} \times \dot{\mathbf{n}} \right)$$

$$= -\gamma \mathbf{m}_0 \cdot \left( \frac{\partial \mathbf{m}_0}{\partial x_i} \times (\mathbf{m} \times \mathbf{H}_{\text{eff}}) \right) - \alpha \mathbf{m}_0 \cdot \left( \frac{\partial \mathbf{m}_0}{\partial x_i} \times (\mathbf{m}_0 \times (\mathbf{v} \cdot \nabla) \mathbf{m}_0) \right) + \tag{9}$$

$$\alpha \mathbf{m}_0 \cdot \left( \frac{\partial \mathbf{m}_0}{\partial x_i} \times (\mathbf{n} \times (-(\mathbf{v} \cdot \nabla) \mathbf{m}_0)) \right) + \alpha \mathbf{m}_0 \cdot \left( \frac{\partial \mathbf{m}_0}{\partial x_i} \times (\mathbf{m} \times \dot{\mathbf{n}}) \right)$$

By defining

$$\mathbf{g} = \left[ \mathbf{m}_0 \cdot \left( \frac{\partial \mathbf{m}_0}{\partial y} \times \frac{\partial \mathbf{m}_0}{\partial z} \right),\ \mathbf{m}_0 \cdot \left( \frac{\partial \mathbf{m}_0}{\partial z} \times \frac{\partial \mathbf{m}_0}{\partial x} \right),\ \mathbf{m}_0 \cdot \left( \frac{\partial \mathbf{m}_0}{\partial x} \times \frac{\partial \mathbf{m}_0}{\partial y} \right) \right] \tag{10}$$

and

$$\ddot{\mathbf{d}} = \begin{bmatrix} d_{11} & d_{12} & d_{13} \\ d_{21} & d_{22} & d_{23} \\ d_{31} & d_{32} & d_{33} \end{bmatrix},\ d_{ij} = \alpha \frac{\partial m_{0k}}{\partial x_i} \frac{\partial m_{0k}}{\partial x_j} \tag{11}$$

Eq. (10) can be further written as

$$\mathbf{g} \times \mathbf{v} + \ddot{\mathbf{d}} \cdot \mathbf{v} = -\gamma \mathbf{m}_0 \cdot \left( \frac{\partial \mathbf{m}_0}{\partial x_i} \times (\mathbf{m} \times \mathbf{H}_{\text{eff}}) \right) - \mathbf{m}_0 \cdot \left( \frac{\partial \mathbf{m}_0}{\partial x_i} \times \dot{\mathbf{n}} \right) +$$

$$\alpha \mathbf{m}_0 \cdot \left( \frac{\partial \mathbf{m}_0}{\partial x_i} \times (\mathbf{n} \times (-(\mathbf{v} \cdot \nabla) \mathbf{m}_0)) \right) + \alpha \mathbf{m}_0 \cdot \left( \frac{\partial \mathbf{m}_0}{\partial x_i} \times (\mathbf{m} \times \dot{\mathbf{n}}) \right) \tag{12}$$

The time averages of those terms related to **n** and **ṅ** in the above equation can be neglected. This leads to

$$\mathbf{g} \times \mathbf{v} + \ddot{\mathbf{d}} \cdot \mathbf{v} = -\left\langle \gamma \mathbf{m}_0 \cdot \left( \frac{\partial \mathbf{m}_0}{\partial x_i} \times (\mathbf{m} \times \mathbf{H}_{eff}) \right) \right\rangle \tag{13}$$

By integrating all the terms of the Eq. (13) with space coordinates, one finally obtains

$$\mathbf{G} \times \mathbf{v} + \vec{\mathbf{D}} \cdot \mathbf{v} = \mathbf{F} \tag{14}$$

The effective force is defined by

$$F_i = \int -\gamma \mathbf{m}_0 \cdot \left( \frac{\partial \mathbf{m}_0}{\partial x_i} \times \langle \mathbf{m} \times \mathbf{H}_{eff} \rangle \right) dV \tag{15}$$

where

$$\langle \mathbf{m} \times \mathbf{H}_{eff} \rangle \approx \langle A \mathbf{n} \times \nabla^2 \mathbf{n} - D \mathbf{n} \times (\nabla \times \mathbf{n}) + \mathbf{n} \times \mathbf{h}(t) \rangle \tag{16}$$

with the first term corresponding to the contribution of magnons. The ratchet motion speed **v** can be analytically obtained by solving Eq. (14).

**Data availability**

The data presented in the current study are available from the corresponding authors on reasonable request.

**Acknowledgements**

The authors gratefully acknowledge the financial support of National Natural Science Foundation of China (Nos. 11972382, 11672339, 11832019), Guangzhou Science and Technology Project (No. 201707020002), and Fundamental Research Funds for the Central Universities.

**Author contributions**

Y.Z. initiated and performed this work and manuscript. W.J.C. conceived and designed the basic idea, structures. W.J.C. performed the simulations and modeling. W.J.C., L.J.L. and Y.Z. analyzed the results and co-wrote the manuscript. All authors contributed to discussion and reviewed the manuscript.

**Competing interests**

The authors declare no competing interests.

**Additional information**

Supplementary information is available for this paper at https://doi.org/xxx.

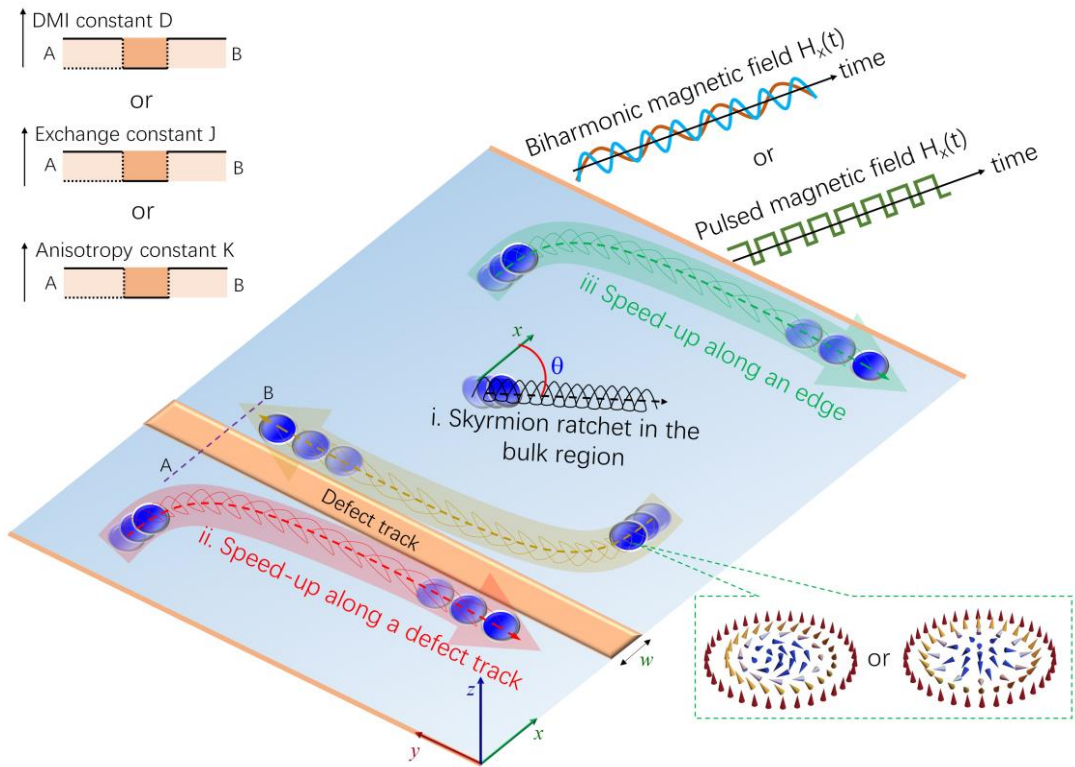

**Fig. 1 Schematic illustrations of the physical model and the basic idea of defect-induced ultrafast ratchet motion of skyrmion under time-varying magnetic fields.** Analogy to other soliton systems, both Bloch and Néel type magnetic skyrmions would be driven into ratchet motions by time-varying forces with temporal asymmetry, e.g., in-plane biharmonic magnetic fields $H_x(t) = H_0[\sin(m\omega t) + \sin(n\omega t + \varphi)]$ with $m$ and $n$ being two coprime integers and $m + n$ being odd, or in-plane pulsed magnetic fields with positive pulses and negative pulses in different magnitudes or widths. In the bulk region (*i.e.*, free of defects or edges), the speed of such ratchet motions is relatively low, though a continuous control of speed and direction of such ratchet motions can be achieved by tuning the field parameters (*e.g.*, the frequency $\omega$, phase $\varphi$ and strength $H_0$ of the biharmonic field, width of the pulsed field, *etc.*). When the skyrmion approaches to a defect track (with a different DMI, exchange or anisotropy from that of the bulk region) or an edge, it will be guided along the defect track/edge and the skyrmion ratchet motion exhibits a remarkable speed-up in comparison with that in the bulk region, due to the remarkably increase of the time-averaged driving force perpendicular to the skyrmion motion when it approaches the defect track/edge. Furthermore, the skyrmion ratchet motion speed and direction along the defect track/edge can be facilely controlled by the field and defect parameters.

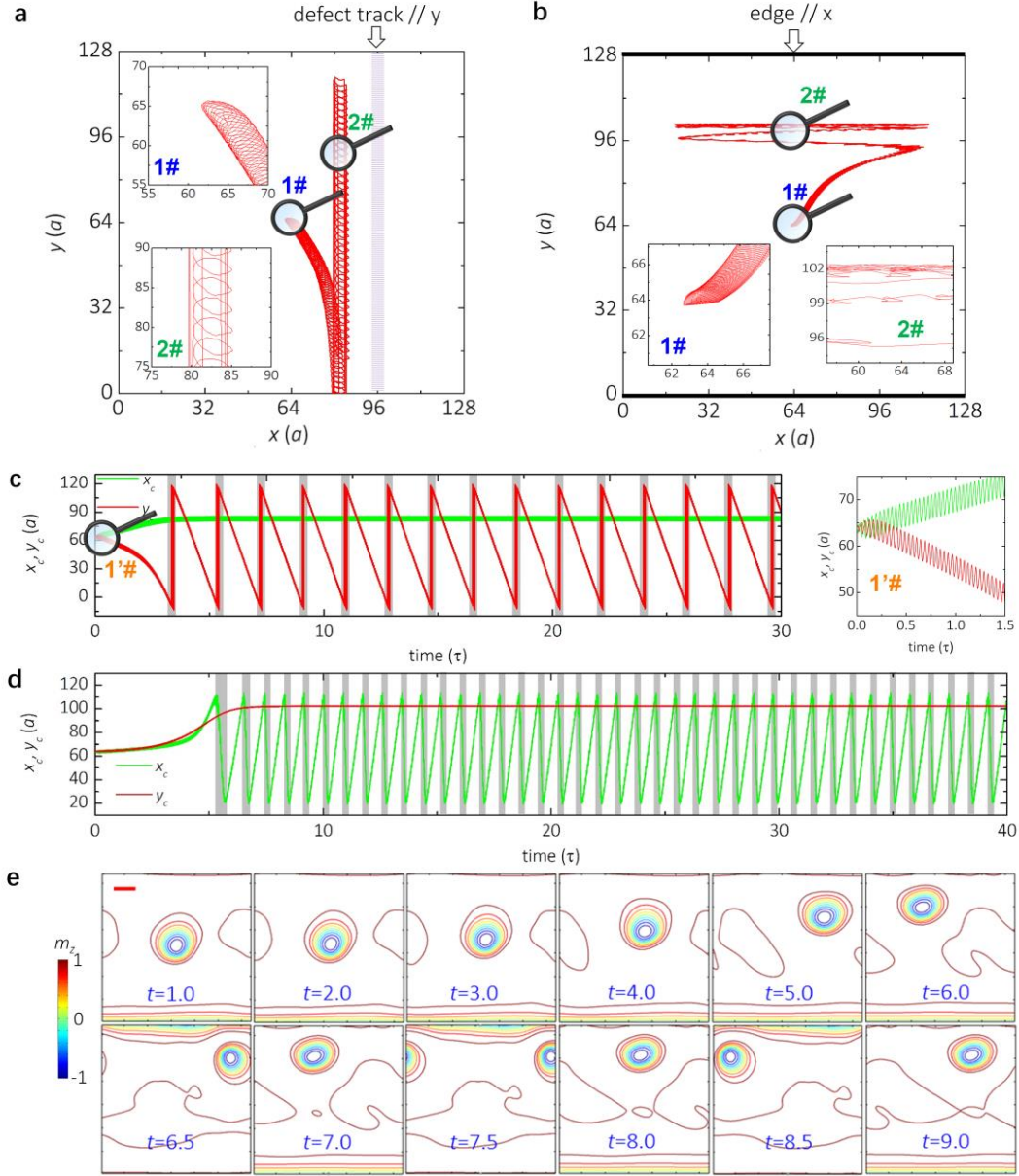

**Fig. 2 Long-time defect-mediated skyrmion ratchet motion under a biharmonic in-plane magnetic field $H_x(t) = 0.003[\sin(\omega t) + \sin(3\omega t)]$. a** Trajectory of a skyrmion moving toward and along a 4$a$-wide defect track ($\eta_{\text{DMI}} = 0.5$) parallel to the $y$-axis (denoted as "defect track // $y$") and locating at $x = 96a$. **b** Trajectory of a skyrmion moving toward and along an edge parallel to the $x$-axis (denoted as "defect track // $x$") and locating at $y = 128a$. To clearly see the speed-up effect, for both cases, the skyrmion is initially located at position (64$a$, 64$a$), which is far enough from the defect track and edge, so that the skyrmion motion therein can be considered as in the bulk region. **c, d** Corresponding long-time evolution curves of the skyrmion position coordinates $x_c$ and $y_c$ of cases **a** and **b**. **e** Snapshots of the contour map of $m_z$ of case **b**, from which the skyrmion configuration can be clearly seen. In **a-d**, the skyrmion position ($x_c$, $y_c$) is determined by $x_c = \int xq dx dy/Q$ and $y_c = \int yq dx dy/Q$, where $q$ is the topological charge density and $Q = \int q dx dy$ is the topological charge. The skyrmion position nearby the boundaries is problematic due to the periodic boundary conditions.

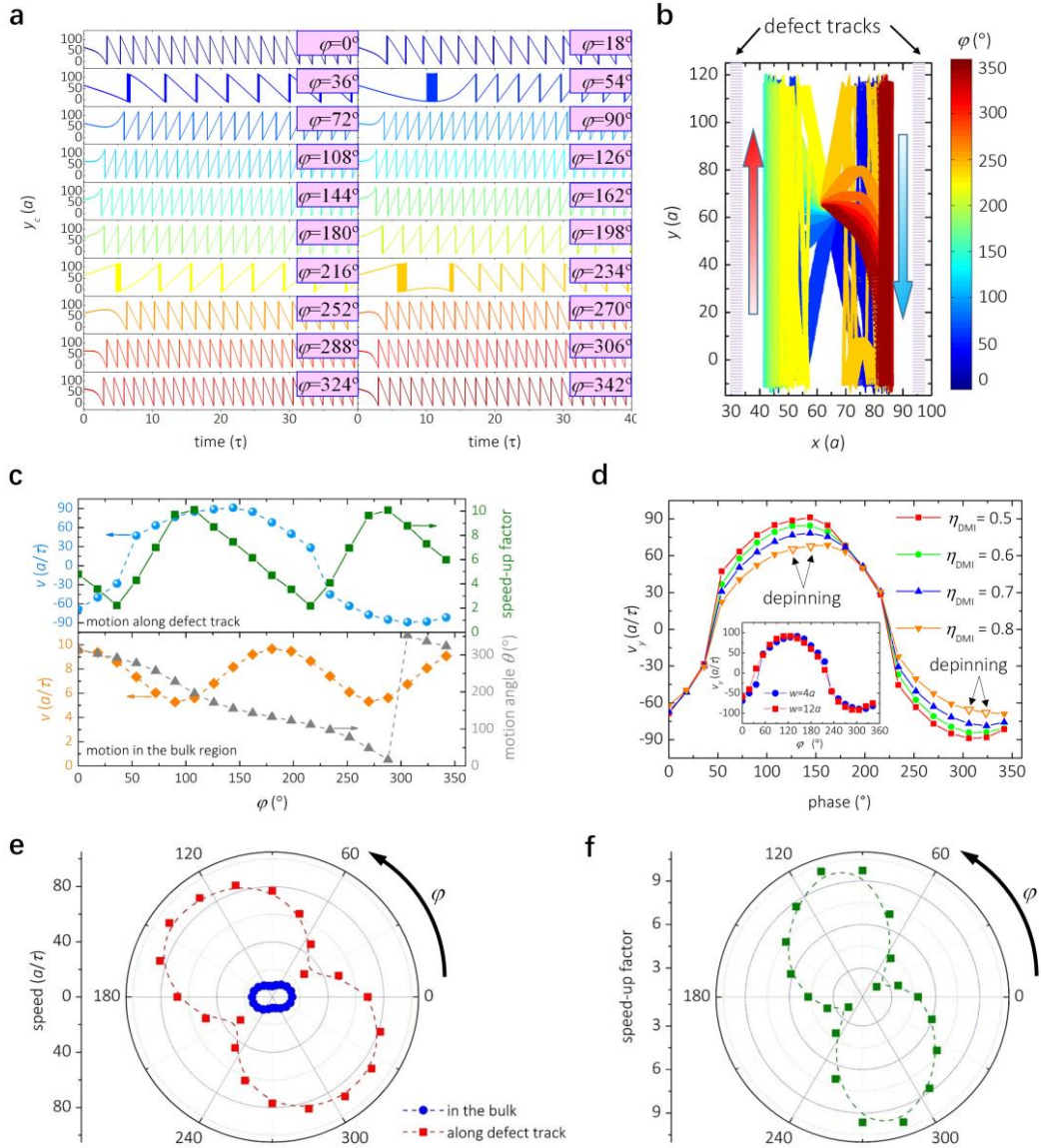

**Fig. 3 The influence of phase $\varphi$ on the skyrmion ratchet motion along defect tracks // y under biharmonic in-plane magnetic fields $H_x(t) = 0.003[\sin(\omega t) + \sin(3\omega t + \varphi)]$.**
**a** Long-time evolution curves of the position coordinate $y_c$ of skyrmion ratchet motion controlled by phase $\varphi$. In the simulation box, two $4a$-wide defect tracks (parallel to the y-axis, $\eta_{DMI} = 0.5$) are located at $x = 32a$ and $x = 96a$. For a given $\varphi$, the skyrmion is initially located at position $(64a, 64a)$ and the skyrmion motion at the beginning can be considered as in the bulk region. Under the driving of $H_x(t)$, the skyrmion will be finally either in upward ratchet motion along the left defect track // y or in downward ratchet motion along the right track // y, depending on the value of phase $\varphi$. **b** Trajectories of skyrmion ratchet motion at various phase $\varphi$. **c** Comparison of the skyrmion ratchet motion velocity along the defect track // y and in the bulk region. In the bulk region, the ratchet motion speed is approximately a sinusoidal function of phase $\varphi$ in period of 180°, with a maximum value of about $9.5a/\tau$. Moreover, the ratchet direction (denoted by the motion angle $\theta$) is rotatable over 360° by changing $\varphi$. For the ratchet motion along the defect track, the motion direction is either upward or downward along the defect track, and the speed is approximately a sinusoidal function of phase $\varphi$ in period of 360° with a maximum value of about $91.3a/\tau$. The speed-up factor is denoted as the ratio of the ratchet motion speed along the defect track to that in the bulk region. It is an oscillating function of phase $\varphi$ in period of 180° with a maximum value of ~10 as shown in the up panel. **d** Skyrmion ratchet speed $v_y$ as a function of phase $\varphi$ of the biharmonic field for the case of "defect track // y" with different defect strength $\eta_{DMI}$. Effect of the width of defect track is shown in the inset. Note, depinning occurs once the force perpendicular to the defect track exceeds the depinning force but the defect strength is not strong enough ($\eta_{DMI} = 0.8$). **e** Skyrmion ratchet motion speed and **f** speed-up factor as a function of phase $\varphi$.

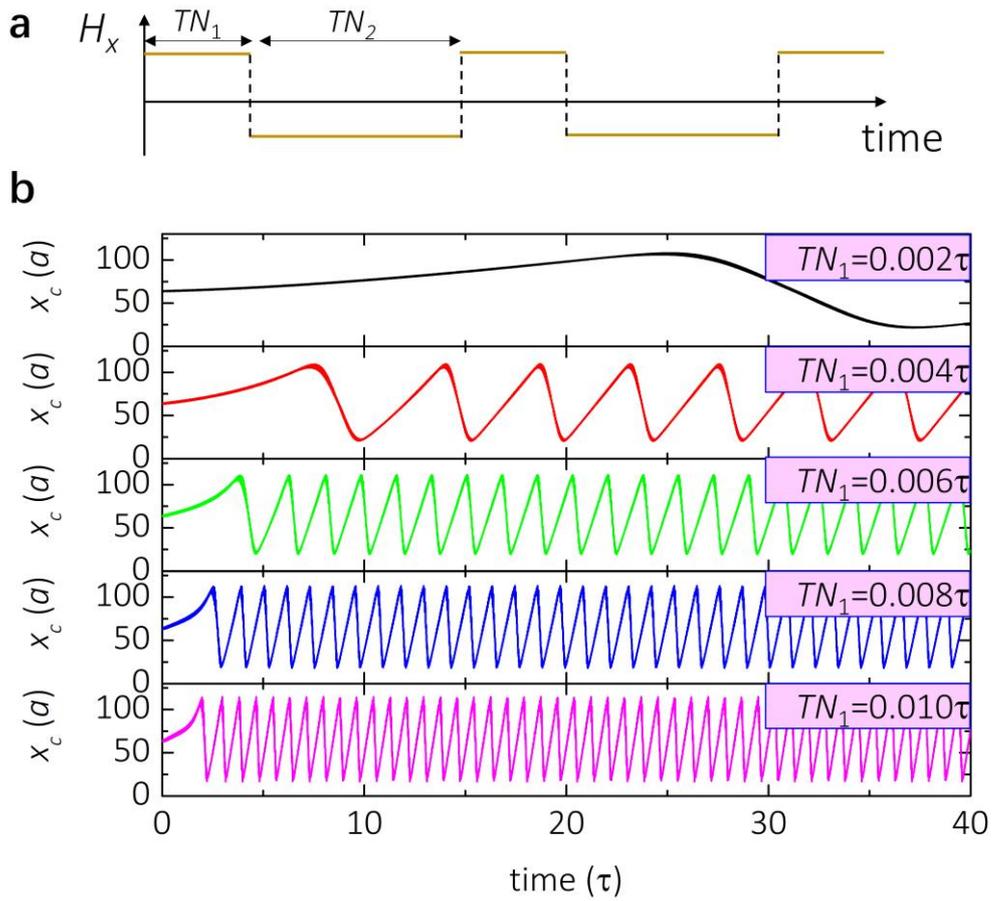

**Fig. 4 Ultrafast skyrmion ratchet motion along an edge // $x$ under pulsed in-plane magnetic fields. a** Schematic of the pulsed fields $H_x(t)$, which are repetitive sequences of alternating positive pulse over a time $TN_1$ and negative pulse over a time $TN_2$. **b** Long-time evolution curves of the position coordinate $x_c$ of skyrmion ratchet motion at various $TN_1$. Here, the magnitudes of the positive and negative pulses are $0.003J/g\mu_B$, and $TN_1 + TN_2 = TN$ is fixed to be 60 ps.

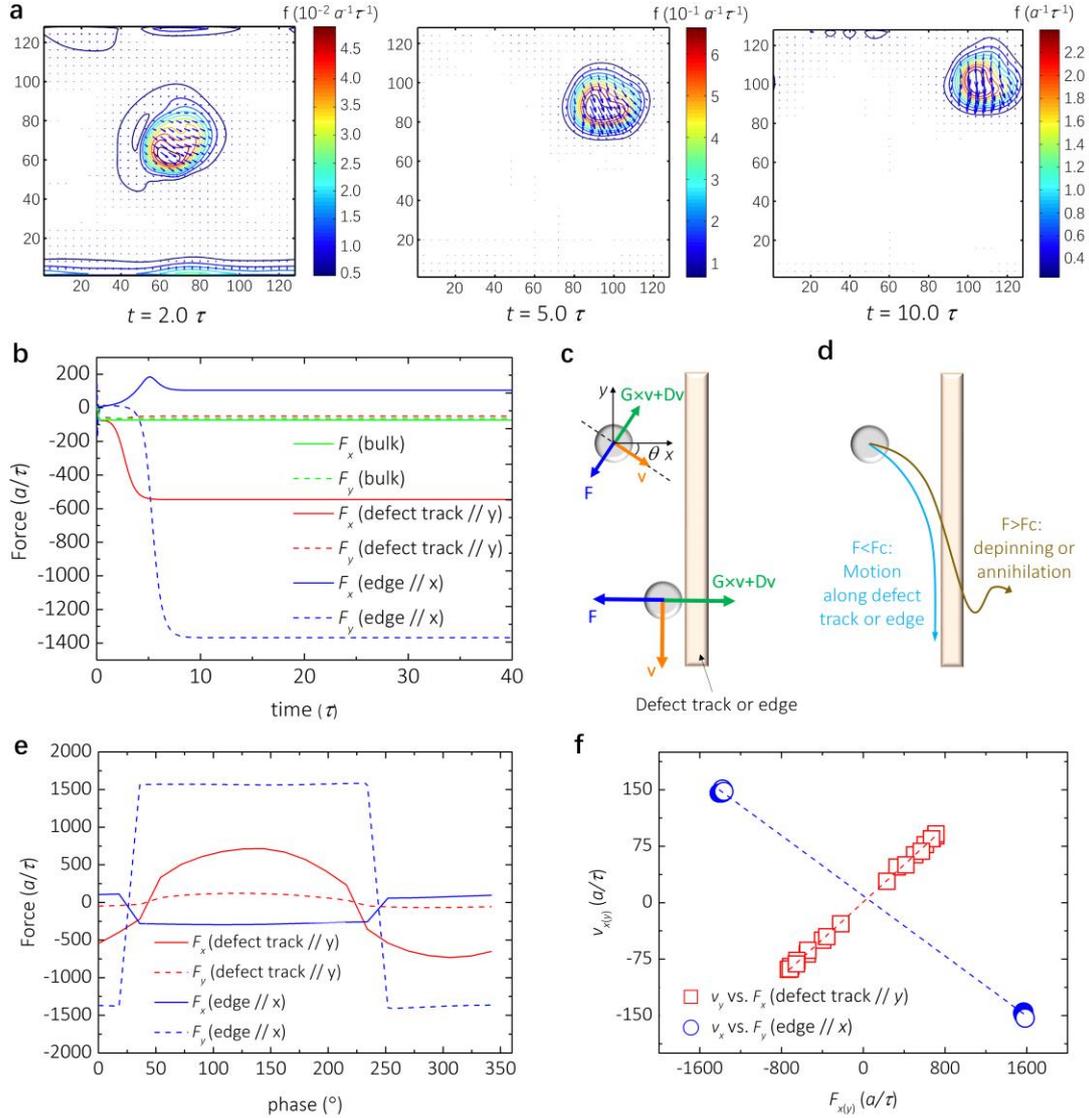

Fig. 5 Force analysis of the defect-induced ultrafast ratchet motion of skyrmion under biharmonic in-plane magnetic fields $H_x(t) = 0.003[\sin(\omega t) + \sin(3\omega t + \varphi)]$. **a** Snapshots of the force density map of skyrmion moving toward and along an edge // $x$. **b** Evolution of time-averaged driving force components $F_{x(y)}$ subjected by skyrmion when it moves in the bulk region (denoted as "bulk"), along a defect track parallel to $y$-axis (denoted as "defect track // $y$"), and along an edge parallel to $x$-axis (denoted as "edge // $x$"). For all the cases, the box size is $128a \times 128a$, and skyrmion is initially located at position $(64a, 64a)$. For the case of "defect track // $y$", the defect track is $4a$-wide with $\eta_{DMI} = 0.5$ and located at $x = 96a$. For the case of "edge // $x$", the edge is located at $y = 128a$. **c** Schematic of the forces subjected by a skyrmion when it is moving in the bulk region and along a defect track or edge. **d** Schematic illustration of two situations when a skyrmion encounters a defect track or edge. Depending on the relative magnitudes of the force perpendicular to the defect track/edge $F$ subjected by skyrmion and the critical force $F_c$ of depinning or annihilation, three situations of skyrmion motion are expected: (i) skyrmion moves along the defect track or edge ($F < F_c$), (ii) passes through the defect track ($F > F_c$ of depinning), and (iii) annihilates at the defect

track or edge ($F > F_c$ of annihilation). **e** Time-averaged driving force components $F_{x(y)}$ subjected by skyrmion as functions of phase $\varphi$ for cases of "defect track // y" and "edge // x". **f** Correlation between the skyrmion ratchet speed $v_{x(y)}$ and force $F_{x(y)}$ for cases of "defect track // y" and "edge // x".

Supplmenetary Information for

# Ultrafast ratchet dynamics of skyrmion by defect engineering under gigahertz magnetic fields

Weijin Chen,[1,2,3] Linjie Liu,[2,3] and Yue Zheng[2,3]★

[1]*School of Materials, Sun Yat-sen University, 510275, Guangzhou, China*
[2]*State Key Laboratory of Optoelectronic Materials and Technologies, Sun Yat-sen University, 510275, Guangzhou, China*
[3]*Micro&Nano Physics and Mechanics Research Laboratory, School of Physics, Sun Yat-sen University, 510275, Guangzhou, China*

★ e-mail: zhengy35@mail.sysu.edu.cn.

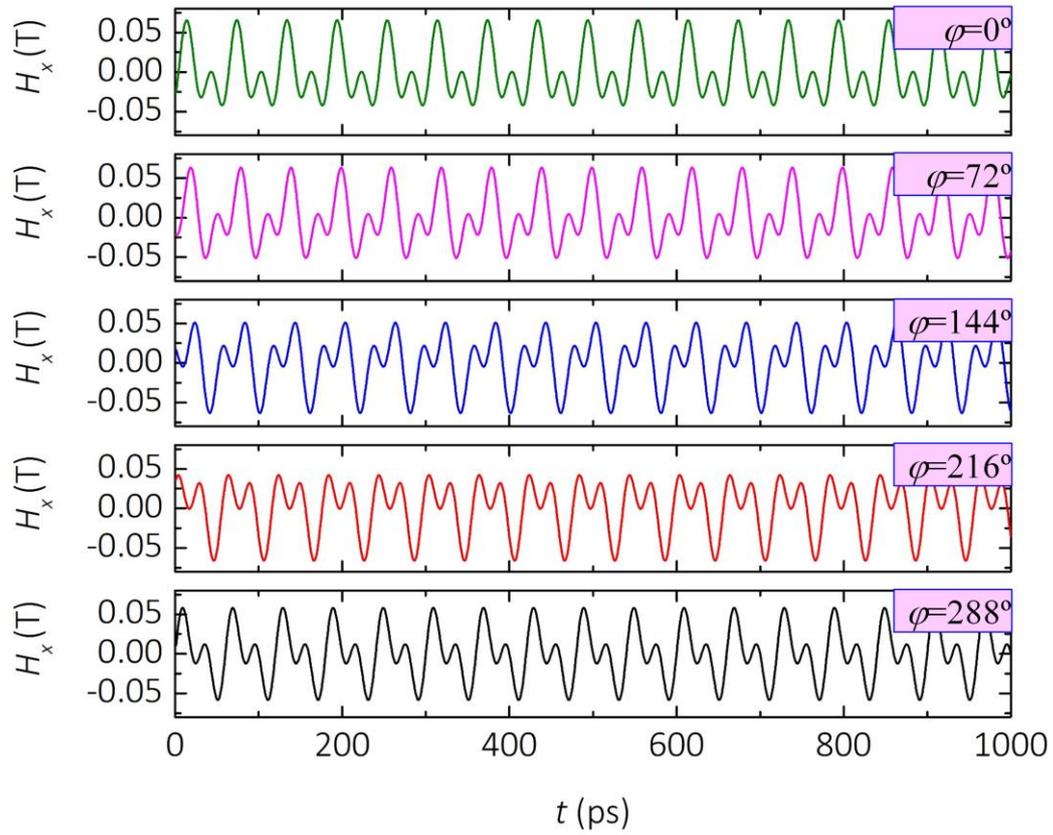

**Fig. S1 The biharmonic in-plane magnetic fields $H_x(t) = H_0[\sin(\omega t) + \sin(3\omega t + \varphi)]$ as functions of time at different $\varphi$.** The angular frequency of the fields $\omega$ is set to be 105 rad/s and $H_0 = 0.003\ J/g\mu_B \approx 0.033$ T.

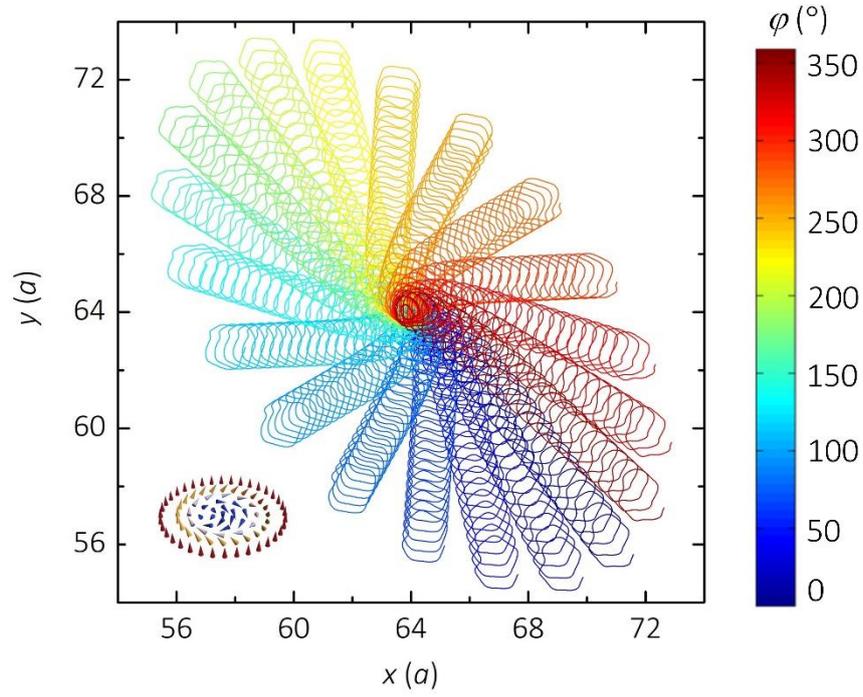

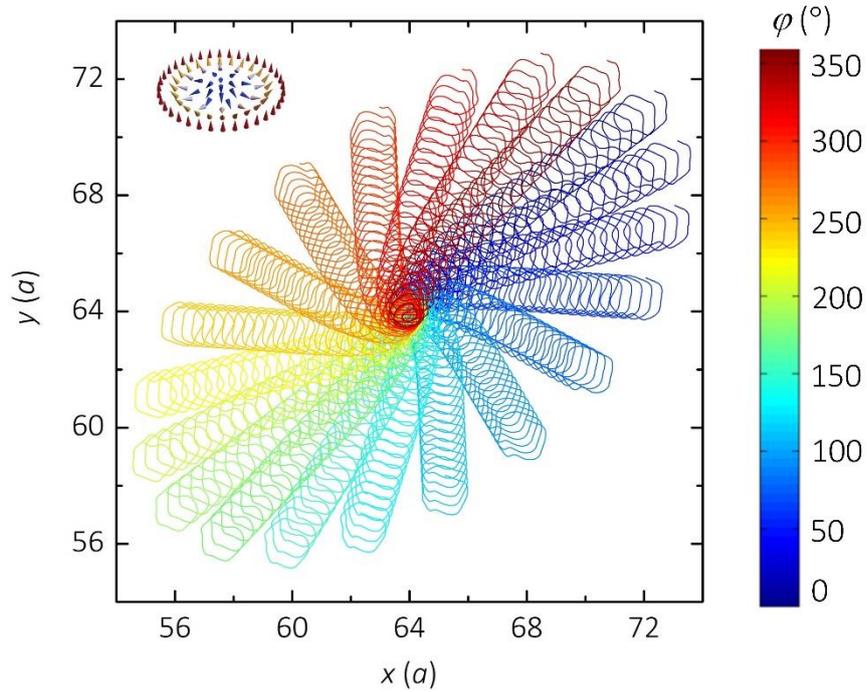

**Fig. S2 Skyrmion ratchet motion under biharmonic in-plane magnetic fields $H_x(t)$ = 0.003[sin($\omega t$) + sin($3\omega t + \varphi$)] at different $\varphi$ in the absence of defects or edges**. **a** Bloch-type skyrmion and **b** Neél-type skyrmion. Trajectories of skyrmion ratchet motion at various phase $\varphi$ are given in different colors. For a given $\varphi$, the skyrmion is initially located at position ($64a$, $64a$).

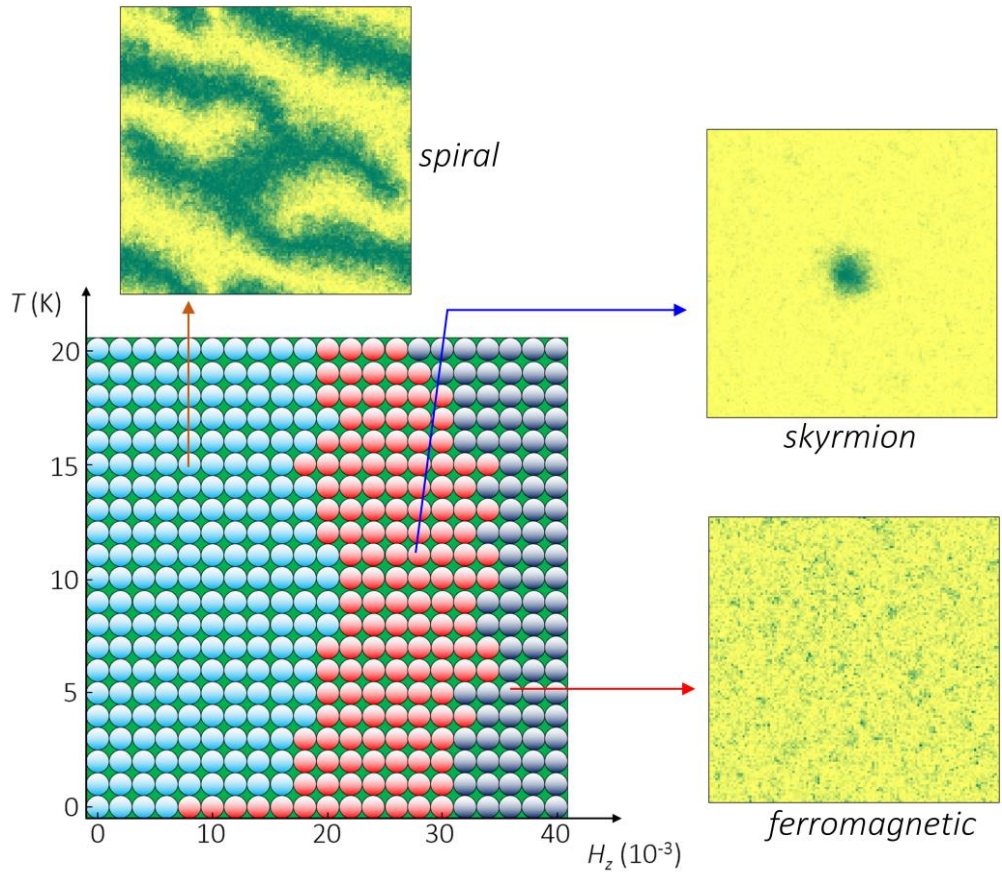

**Fig. S3 Temperature-field *T*-$H_z$ phase diagram of single-skyrmion state**. The phase diagram is obtained by performing LLG simulations at various temperature *T* and out-of-plane magnetic field $H_z$ on a 128 × 128 spin lattice with a single-skyrmion state being the trivial state. At a given temperature, the single-skyrmion state can be stabilized at an intermediate range of $H_z$. At smaller $H_z$, the spiral state is more stable. At larger $H_z$, the ferromagnetic state is more stable.

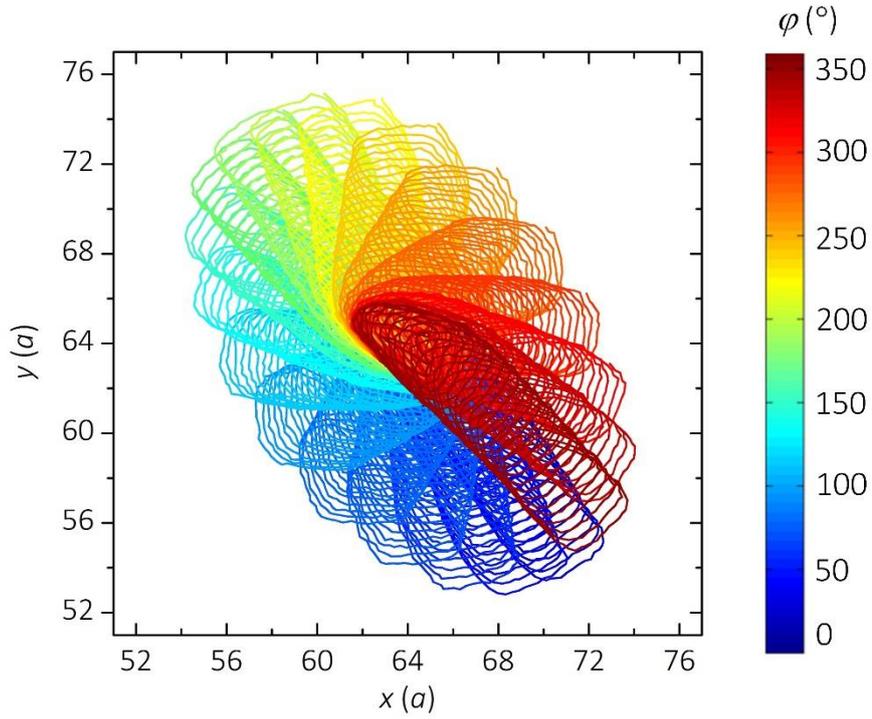

**Fig. S4 Skyrmion ratchet motion in the bulk region in system with double defect tracks under biharmonic in-plane magnetic fields $H_x(t) = 0.003[\sin(\omega t) + \sin(3\omega t + \varphi)]$.** Trajectories of skyrmion ratchet motion at various phase $\varphi$ are given in different colors. In the simulation box, two $4a$-wide defect tracks (parallel to the $y$-axis, $\eta_{DMI} = 0.5$) are located at $x = 32a$ and $x = 96a$. For a given $\varphi$, the skyrmion is initially located at position ($64a$, $64a$). This result corresponds to Fig. 3 in the main text.

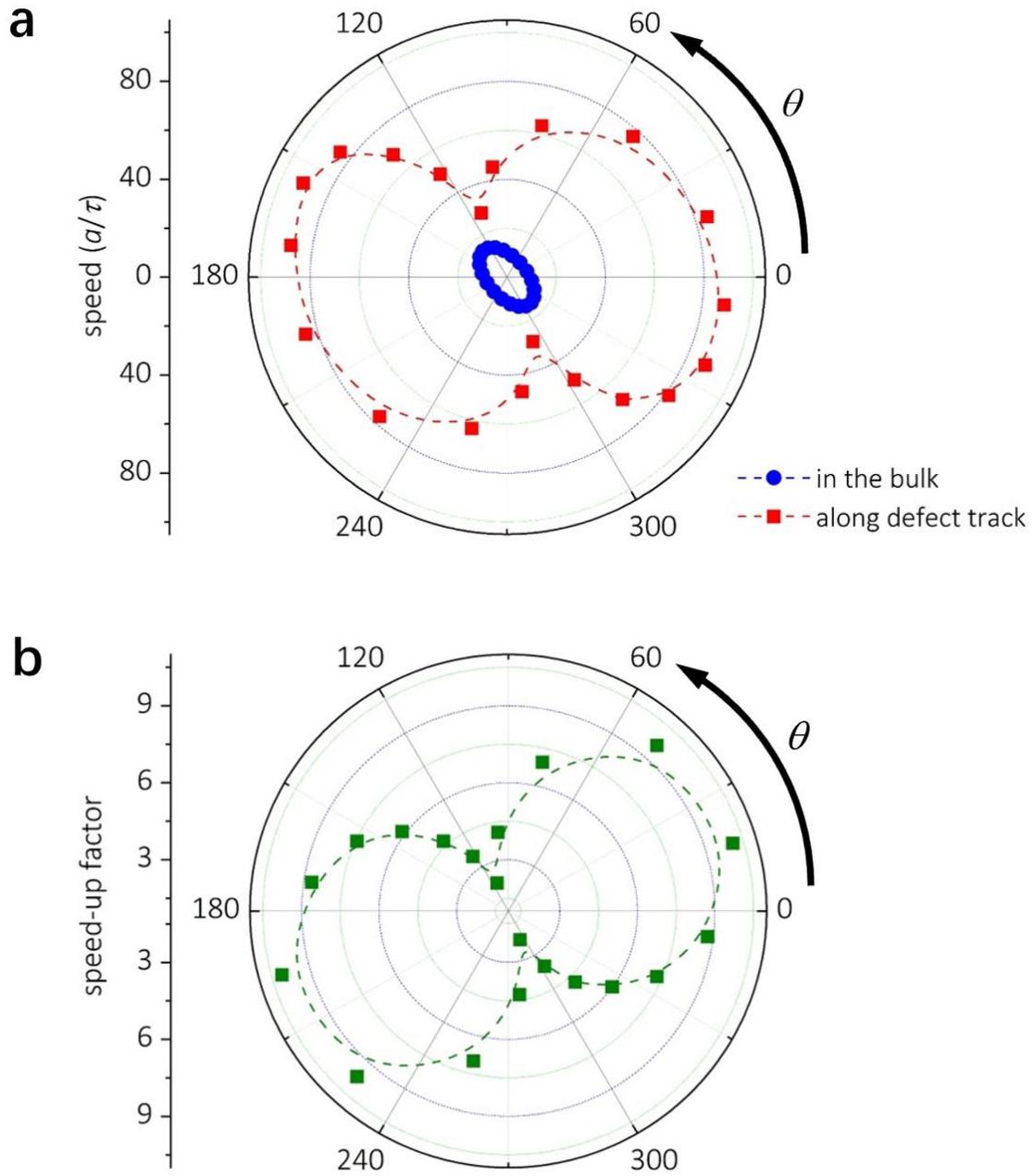

**Fig. S5 Polar plots of the skyrmion ratchet motion speed and speed-up factor along defect tracks // y under biharmonic in-plane magnetic fields $H_x(t) = 0.003[\sin(\omega t) + \sin(3\omega t + \varphi)]$. a** Skyrmion ratchet motion speed and **b** speed-up factor as a function of skyrmion motion direction $\theta$ in the bulk region. This result corresponds to Fig. 3 in the main text.

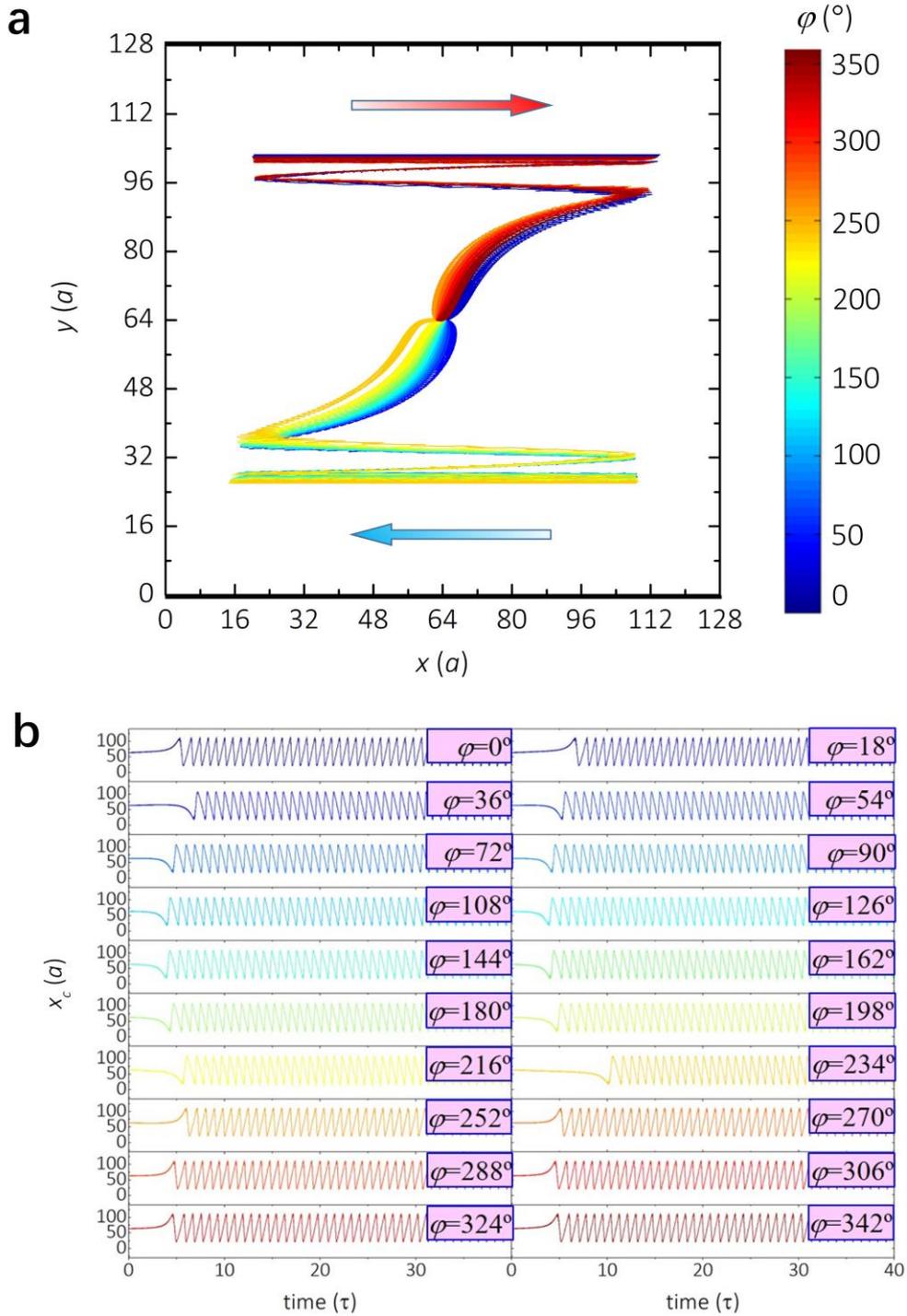

**Fig. S6 Ultrafast skyrmion ratchet motion along edges // $x$ under biharmonic in-plane magnetic fields $H_x(t) = 0.003[\sin(\omega t) + \sin(3\omega t + \varphi)]$. a** Trajectories of skyrmion ratchet motion at various phase $\varphi$ are plotted in different colors. **b** Long-time evolution curves of the position coordinate $x_c$ of skyrmion ratchet motion at various phase $\varphi$. The simulation box size is $128a \times 128a$, periodic along $x$ direction and with two edges at $y = 0a$ and $y = 128a$. For a given phase $\varphi$, skyrmion is initially located at position ($64a$, $64a$).

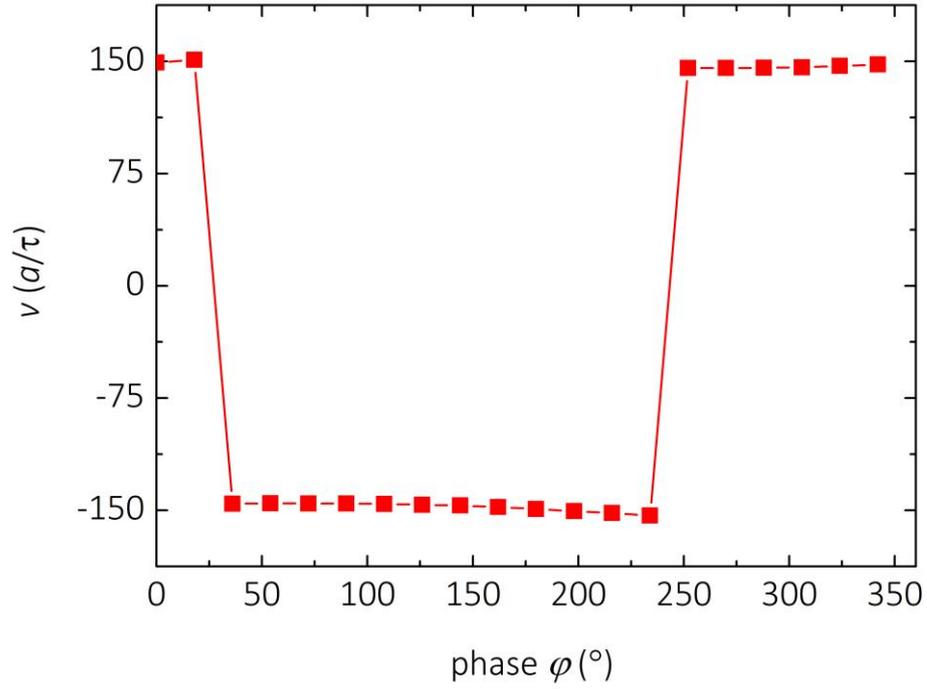

**Fig. S7 Skyrmion ratchet speed $v$ along edges // $x$ as a function of phase $\varphi$ of the biharmonic field $H_x(t) = 0.003[\sin(\omega t) + \sin(3\omega t + \varphi)]$.** This result corresponds to Fig. S6.

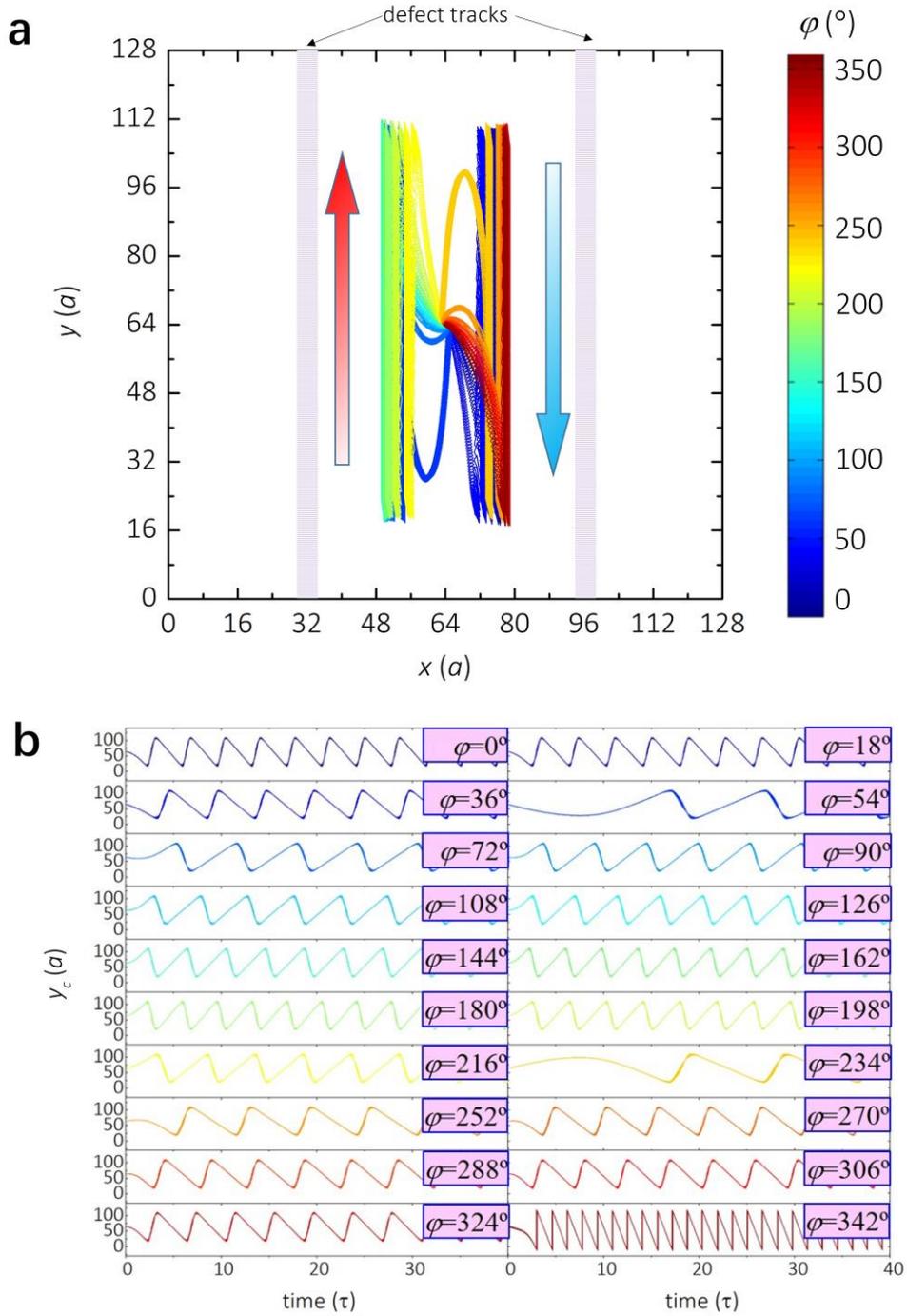

**Fig. S8 Skyrmion ratchet motion along defect tracks // y with a different exchange interaction from that of the bulk region under biharmonic in-plane magnetic fields $H_x(t) = 0.003[\sin(\omega t) + \sin(3\omega t + \varphi)]$. a** Trajectories of skyrmion ratchet motion at various phase $\varphi$ are plotted in different colors. **b** Long-time evolution curves of the position coordinate $y_c$ of skyrmion ratchet motion at various phase $\varphi$. In the simulation box, two $4a$-wide defect tracks (parallel to the $y$-axis) are located at $x = 32a$ and $x = 96a$. The exchange interaction constant in the defect tracks is set to be half of that in the bulk region. For a given $\varphi$, the skyrmion is initially located at position ($64a$, $64a$) and the skyrmion motion at the beginning can be considered as in the bulk region.

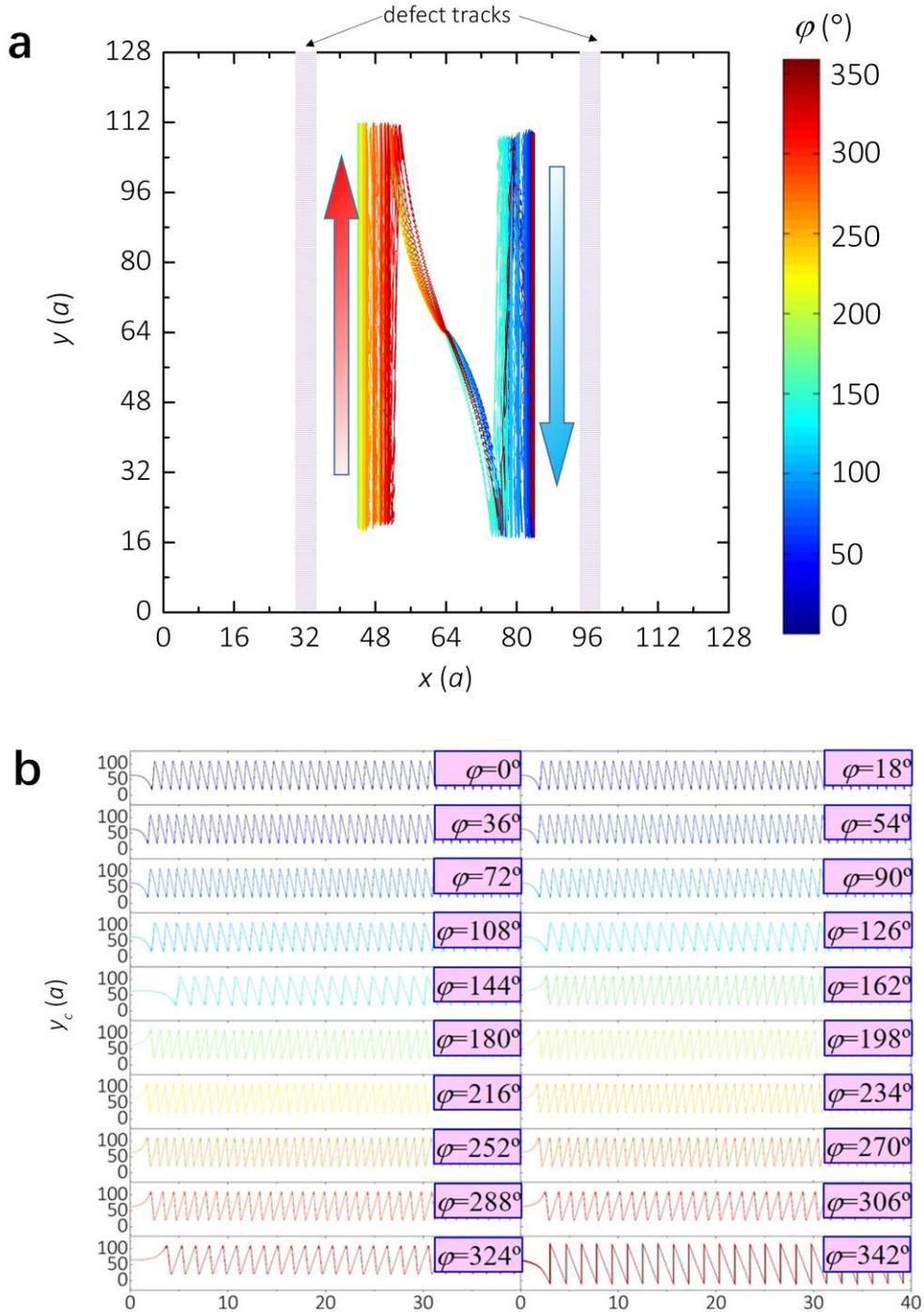

**Fig. S9 Skyrmion ratchet motion along defect tracks // y with a different anisotropy from that of the bulk region under biharmonic in-plane magnetic fields $H_x(t) = 0.003[\sin(\omega t) + \sin(3\omega t + \varphi)]$. a** Trajectories of skyrmion ratchet motion at various phase $\varphi$ are plotted in different colors. **b** Long-time evolution curves of the position coordinate $y_c$ of skyrmion ratchet motion at various phase $\varphi$. In the simulation box, two $4a$-wide defect tracks (parallel to the $y$-axis) are located at $x = 32a$ and $x = 96a$. The anisotropy constant is set to be $K=0.015$ in the defect tracks and $K=0$ elsewhere. For a given $\varphi$, the skyrmion is initially located at position $(64a, 64a)$ and the skyrmion motion at the beginning can be considered as in the bulk region.

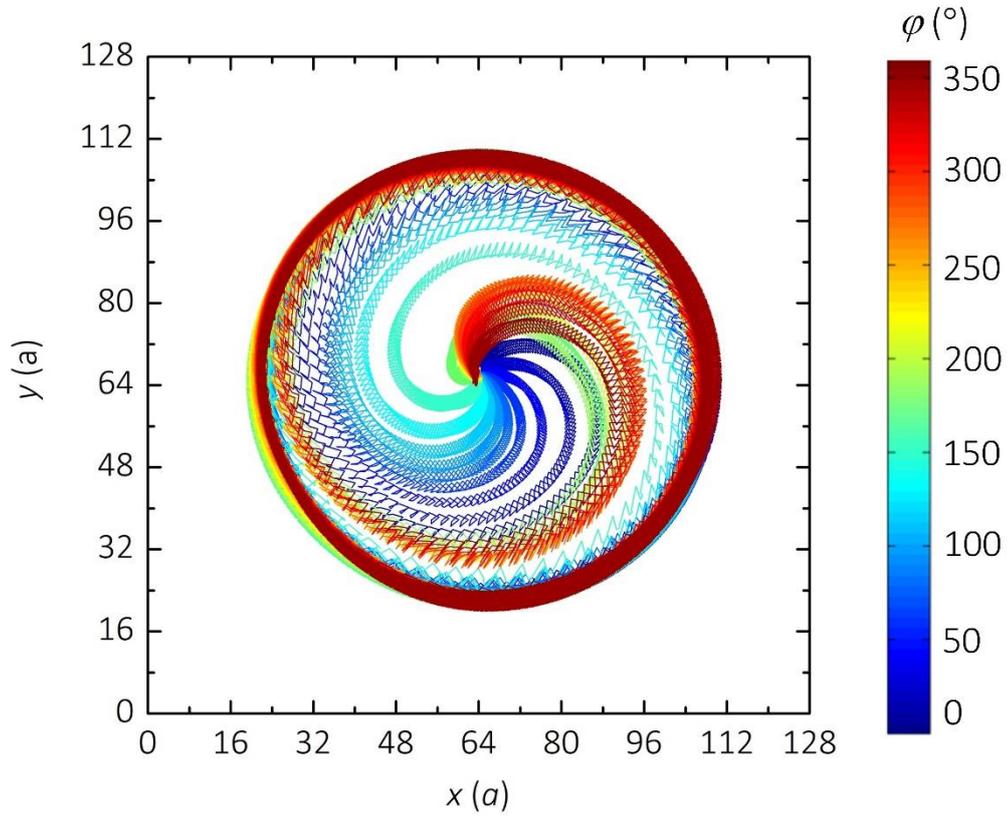

**Fig. S10 Ultrafast skyrmion ratchet motion in a circular disk under biharmonic in-plane magnetic fields $H_x(t) = 0.003[\sin(\omega t)+\sin(3\omega t + \varphi)]$.** Trajectories of skyrmion ratchet motion at various phase $\varphi$ are given in different colors. The radius of the disk is equal to $64a$. For a given $\varphi$, the skyrmion is initially located at the center. Under the driving force of biharmonic field, the skyrmion finally approaches the edge of the disk with a remarkable angular speed over 100 rad/s.

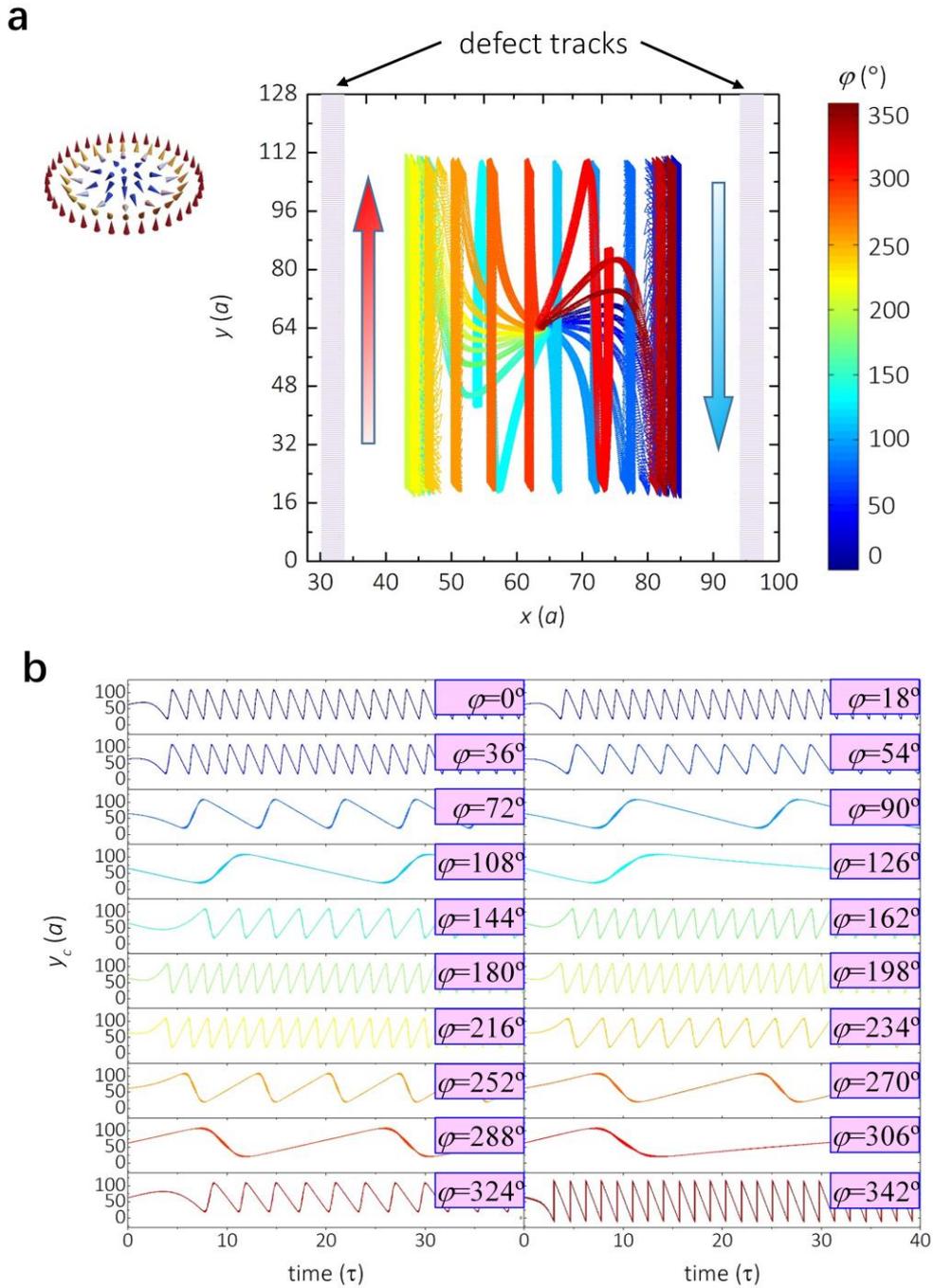

**Fig. S11 Ultrafast ratchet motion of Neél-type skyrmion along defect tracks // y under biharmonic in-plane magnetic fields $H_x(t) = 0.003[\sin(\omega t) + \sin(3\omega t + \varphi)]$. a** Trajectories of skyrmion ratchet motion at various phase $\varphi$ are plotted in different colors. **b** Long-time evolution curves of the position coordinate $y_c$ of skyrmion ratchet motion at various phase $\varphi$. In the simulation box, two $4a$-wide defect tracks (parallel to the $y$-axis, $\eta_{DMI} = 0.5$) are located at $x = 32a$ and $x = 96a$. For a given $\varphi$, the skyrmion is initially located at position $(64a, 64a)$.